\documentclass{IEEEtran}
\usepackage{amsmath} % assumes amsmath package installed
\usepackage{amssymb}  % assumes amsmath package installed
\usepackage[english]{babel}
\usepackage{lipsum}
\usepackage{nicematrix}
\usepackage{graphicx}
\usepackage{amsmath}

\usepackage{textcomp}
\usepackage{amssymb}
\usepackage{bm}
\usepackage{mathtools}

\usepackage{comment}
\usepackage{textcomp}
\usepackage{microtype}

\usepackage{float}
\usepackage{booktabs,makecell,tabularx}

\newcolumntype{C}[1]{>{\centering\arraybackslash}p{#1}}
\newcolumntype{L}{>{\raggedright\arraybackslash}X}

\usepackage{siunitx}
\usepackage{booktabs}
\usepackage{multirow}
\usepackage{algorithm}
\usepackage{algpseudocode}
\usepackage{etoolbox}

\makeatletter
\newcommand*{\algrule}[1][\algorithmicindent]{%
  \makebox[#1][l]{%
    \hspace*{.2em}% <------------- This is where the rule starts from
    \vrule height .75\baselineskip depth .25\baselineskip
  }
}

\newcount\ALG@printindent@tempcnta
\def\ALG@printindent{%
    \ifnum \theALG@nested>0% is there anything to print
    \ifx\ALG@text\ALG@x@notext% is this an end group without any text?
    \else
    \unskip
    \ALG@printindent@tempcnta=1
    \loop
    \algrule[\csname ALG@ind@\the\ALG@printindent@tempcnta\endcsname]%
    \advance \ALG@printindent@tempcnta 1
    \ifnum \ALG@printindent@tempcnta<\numexpr\theALG@nested+1\relax
    \repeat
    \fi
    \fi
}
\patchcmd{\ALG@doentity}{\noindent\hskip\ALG@tlm}{\ALG@printindent}{}{\errmessage{failed to patch}}
\patchcmd{\ALG@doentity}{\item[]\nointerlineskip}{}{}{} % no spurious vertical space
\ifCLASSINFOpdf
\else
\fi
\ifCLASSOPTIONcompsoc
 \usepackage[caption=false,font=normalsize,labelfont=sf,textfont=sf]{subfig}
\else

 \usepackage[caption=false,font=footnotesize]{subfig}
\fi
\usepackage[english]{babel}
\usepackage{lipsum}
\usepackage{comment}
\usepackage{nicematrix}
\usepackage{graphicx}
\usepackage{amsmath}

\begin{document}

\title{Nonlinear Model Predictive Control for Navy Microgrids with Stabilizing Terminal Ingredients}

\author{\IEEEauthorblockN{\textsuperscript{}  Saskia Putri, Xiaoyu Ge, Faegheh Moazeni, Javad Khazaei$^*$}\\
\IEEEauthorblockA{\textit{Rossin college of engineering and applied science} \\
\textit{Lehigh University},
Bethlehem PA, USA \\
Emails: \textit{sap322@lehigh.edu, xig620@lehigh.edu,moazeni@lehigh.edu, khazaei@lehigh.edu}}
% \\
% \color{blue}This work has been submitted to the IEEE for possible publication. Copyright may be transferred without notice, after which this version may no longer be accessible.
\thanks{This research was in part under support from the Department of Defense, Office of Naval Research award number N00014-23-1-2602.}
}
\IEEEpubidadjcol
\IEEEpubid{\begin{minipage}{\textwidth}\ \\[25pt] \centering
  \color{blue}This work has been submitted to the IEEE for possible publication. Copyright may be transferred without notice, after which this version may no longer be accessible.
\end{minipage}}

\maketitle
% Diverging from conventional linear models, this method accounts for the nonlinear dynamics of MGs as the prediction model, employing a streamlined representation of MVDC dynamics.
\begin{abstract}
This paper presents a novel control strategy for medium voltage DC (MVDC) naval shipboard microgrids (MGs), employing a nonlinear model predictive controller (NMPC) enhanced with stabilizing features and an intricate droop control architecture. This combination quickly regulates the output voltage and adeptly allocates supercapacitors for pulsed power loads (PPLs), while the battery energy storage system (BESS) and auxiliary generators handle the steady state loads. A key feature of this study is the formulation of terminal cost and constraints, providing recursive feasibility and closed-loop stability in the Lyapunov sense, that offers a more robust and effective approach to naval power and energy management. By comparing the proposed Lyapunov-based NMPC with conventional PI controller under fluctuating PPLs, the control robustness is validated. 
% The findings highlight the proposed control algorithm’s capability to ensure voltage stability and effective power distribution, critical for the success of naval missions.
\end{abstract}
\begin{IEEEkeywords}
Nonlinear Control, Medium Voltage DC, Naval Power and Energy Systems, Power Sharing, Voltage restoration, Lyapunov Equation.
\end{IEEEkeywords}

% \fi
\IEEEpeerreviewmaketitle

\section{Introduction}
Medium voltage DC (MVDC) SPS enables lower transmission loss, harmonics and interference minimization, and potential weight reduction for shipboard power systems (SPSs) compared to AC systems \cite{faddel2019coordination}. However, MVDC systems in naval vessels face issues pertaining to voltage variation and power quality degradation due to PPLs and continuous variation in propulsion speed. Therefore, the optimal control strategy of the MVDC SPS is imperative \cite{nguyen2021energy}. 

To fully incorporate the complicated nature of SPS, a control approach capable of handling multi-variable, nonlinear, and multi-objective models, system constraints, and disturbances is required. Consequently, model predictive control (MPC) constitutes a better control option that explicitly considers input-output constraints to incorporate voltage deviation and maximum power dispatch from the generators \cite{bordons_model_2020}. Various studies have delved into utilizing MPC for the energy management of  MVDC naval shipboard MGs. This includes optimizing power distribution based on load priorities \cite{zohrabi2017reconfiguration}, maintaining voltage stability and proper power sharing among energy sources \cite{saad2019small}, accommodating the challenges of warship power systems, including MVDC and MVAC and hybrid MVAC/MVDC \cite{young2023model}.

Despite the aforementioned advantages, relying solely on the classical formulation of MPC in the Navy SPS may fail to deliver stable closed-loop systems. This is because classical MPC employs a finite horizon that is unable to deliver asymptotic stability \cite{chen1998quasi}. To address this, a suitable tuning of the MPC design parameters can be utilized by adding terminal ingredients. The terminal states are penalized such that the terminal cost constraints the infinite horizon cost of the nonlinear system controlled by a ``fictitious" local linear state feedback \cite{chen1998quasi}. The terminal costs are calculated as a Lyapunov function for a linear state feedback controller that is locally stabilizing \cite{lazar2018computation} such that the states of the system always return to the origin. Investigation of Navy SPS stability, utilizing Lyapunov stability theorem was studied in \cite{hosseinzadehtaher2020self} for self-healing battery system and in \cite{zohrabi2019efficient} to ensure voltage performance within the safety limits. However, none of the existing MPC approaches considered power sharing between multiple energy resources, including batteries, supercapacitors, and generators within a ship. There is currently a research gap for energy management system of a navy ship that offers closed-loop stability, regulate the DC voltage on the ship, and accommodate multiple energy resources in the presence of PPLs and CPLs.
 
To address the existing gaps, this paper proposes a stability-guaranteed nonlinear MPC (NMPC) combined with complex droop control for power sharing among multiple resources and for voltage restoration of the proposed MVDC naval shipboard microgrids (MGs). The main contributions include:
\begin{enumerate}
   \item Development of nonlinear MPC control algorithms for MVDC naval shipboard MGs with guaranteed stability and recursive feasibility through the addition of terminal ingredients. 
   \item Integration of a reduced order model of MVDC naval shipboard MGs with virtual capacitive and resistive droop controllers as the prediction model to maintain voltage restoration, power balance, and obtain optimal power sharing among the available energy sources. 
   \item Investigation of the integrated control algorithms (complex droop control and Lyapunov-based NMPC) against fluctuation in PPL both in duration and magnitude as well as noise-influenced load uncertainty and compared to conventional PI controller. 
\end{enumerate}
The rest of the paper is structured as follows. Mathematical modeling of the navy shipboard power system is detailed in Section II. The problem formulation of the stability-guaranteed nonlinear MPC is described in Section III. Section IV illustrates the validation of the proposed control algorithm through a variety of case studies. Section V concludes the paper.
\begin{figure}[t!]
    \vspace{-0.5cm}
    \centering
    \includegraphics[width =1\columnwidth]{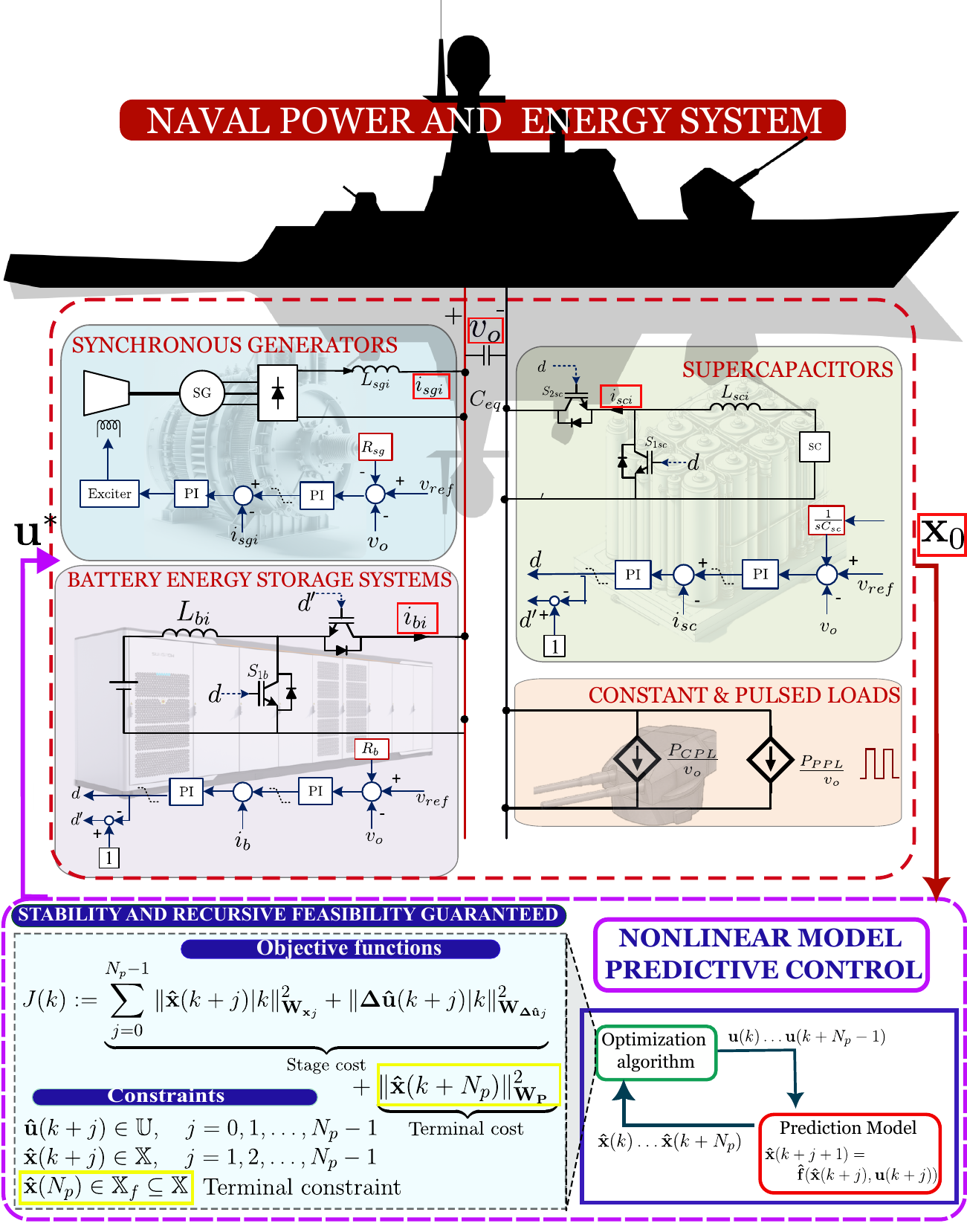}
    \caption{Proposed control scheme for hybrid DC navy MGs}
    \vspace{-0.5cm}
    \label{fig:mvdclayout}
\end{figure}

\section{MVDC NAVAL SHIPBOARD MICROGRIDS}
\subsection{Structure of the proposed MVDC system}
The proposed control scheme for an MVDC naval shipboard MGs is illustrated in Fig.~\ref{fig:mvdclayout}. This system integrates synchronous generators (SGs), battery energy storage systems (BESSs), and supercapacitors (SCs). The paper's objective is to ensure a voltage regulation and equal power sharing among the distributed generation units (DGUs). To achieve this goal, a two-level control scheme is proposed: the primary control to ensure equal power sharing among DGUs to provide a dynamic power sharing,  and a secondary control for voltage regulation and stable operation, employing a Lyapunov-based NMPC. 
\subsection{Mathematical model of MVDC naval shipboard MGs}
As DC power systems often operate over short distances, an equivalent circuit of the system can be utilized expressed as follows \cite{sulligoi2014multiconverter}:
\vspace{-0.1cm}
\begin{subequations} \label{dynmvdc}
        \begin{align}
        C_{\text{eq}} \dot{V_o} &= \frac{}{}\sum_{i \in \mathcal{N}_{SG}} (I_{\text{SG}i}) +\sum_{i \in \mathcal{N}_{B}} (I_{\text{B}i}) +\sum_{i \in \mathcal{N}_{SC}} (I_{\text{SC}i}) \nonumber \\
        &- \frac{P_{\text{CPL}}}{V_o} -\frac{P_{\text{PPL}}}{V_o} \label{eqn:cap_voltage} \\
        L_{i} \dot{I}_{i} &= V_{\text{ref}} - R_{i}I_{i} - V_o + \delta V \quad \forall i \in \{\mathcal{N}_{SG},\mathcal{N}_{B}\} \label{eqn:sg_current} \\
        L_{i}\dot{I}_{i} &= V_{\text{ref}} - R_{i}I_{i} - V_{Ci} - V_o \quad \forall i \in \mathcal{N}_{SC} \label{eqn:sc_current} \\
        C_{i}\dot{V}_{Ci} &= I_{i} \quad \forall i \in \mathcal{N}_{SC} \label{eqn:sc_voltage}
         % C_{m} \dot{SOC}_{i} & = -V_oI_{i}\eta_i \quad \forall i \in \mathcal{N}_{ESS}\label{eqn:SOC}
        \end{align}
\end{subequations}
where $C_{eq}$ is the equivalent capacitor in $F$, $L_i$ is the inductance of the $i$-th generation unit in $H$, $V_o$ denotes the output voltage in $V$, $I_i$ as the current of the $i$-th generation unit in $A$, $R_{i}$, $C_{SCi}$ are the resistive droop gains for the conventional generators and BESSs, and capacitive droop gains for SCs, respectively. Details on the complex droop control design for the ship and the parameters can be found in \cite{hosseinipour2023multifunctional}. To ensure that the SCs are operated during transient conditions and deactivated during steady state, capacitive droop characteristics and voltages of the SCs are embedded in \eqref{eqn:sc_current}, with additional SCs' voltage dynamics in \eqref{eqn:sc_voltage} adopted from \cite{hosseinipour2023multifunctional}. Furthermore, $\delta V$ is added to the system's dynamics to compensate for the virtual impedance ($\delta V \approx R_iI_{i}$) such that constant MVDC bus voltage can be obtained. Furthermore, several corresponding hard constraints must be added to the system's variables as a result of the limitation in physical rules. In this study, output voltage is limited within $\pm 5\%$ of 6kV, and each generation units' output power are limited in MW.

\section{Nonlinear Model Predictive Control with Stabilizing Terminal Ingredients}
This section presents the strategy of the proposed NMPC algorithm with stabilizing terminal ingredients, leveraging on the Lyapunov equation, herein referred to as Lyapunov-based NMPC (LNMPC). The main goal is to ensure voltage restoration and a stable trajectory with proper load sharing among the available generation units in the MVDC naval shipboard MGs under mission loads. Given the system's dynamic functions in \eqref{dynmvdc}, the nonlinear state space representation is expressed as:
\vspace{-0.1cm}
\begin{equation}
\vspace{-0.1cm}
    \dot{\bm{x}} = \mathbf{f}(\bm{x}(t),\bm{u}(t),\bm{d}(t)), \quad \bm{x}(0) = \bm{x}_0 \label{ssmvdc}
\end{equation}
where $u  \in \mathbb{R} = \delta V$ denotes the control input, $\bm{x} \in \mathbb{R}^{n_x} = [V_o, \ I_{SGi} \ I_{Bi} \ I_{SCi} \ VC_{SCi}]^T$ $ \forall i = \{1,2\}$ is the state variable vector, with $\bm{x}_0$ represents the initial values of the state variables, and $\bm{d} \in \mathbb{R}^{n_d} = [P_{CPL} \ P_{PPL}]^T$ represents the disturbances of the system. Integrating the continuous state space model in \eqref{ssmvdc} with zero-order-hold method \cite{franklin1998digital} over a sampling period $\delta t = 5e^{-3}s$, discrete time representation of the system's dynamics is expressed as follows: 
\vspace{-0.1cm}
\begin{align}
\vspace{-0.1cm}
    \bm{x}(k+1) &= \mathbf{f}_d(\bm{x}(k),\bm{u}(k),\bm{d}(k)) \nonumber \\
    &= \bm{x}(k) + \int_{k}^{k + \delta t} \mathbf{f}(\bm{x}(\tau),\bm{u}(\tau),\bm{d}(\tau)) d\tau
 \label{disdyn}
\end{align}
\vspace{-0.5cm}

Furthermore, the prediction model can be developed utilizing the discretized state space model in~\eqref{disdyn} over the prediction horizon ($N_p  \in \mathbb{N}^+ $), expressed as follows:
\begin{align}
    \bm{x}(k+j+1) &= \mathbf{f_d}(\bm{x}(k+j),\bm{u}(k+j),\bm{d}(k+j)) \nonumber \\ & \quad \quad \forall j \in \{0,\dots,N_p-1\} \label{nmpc_pred}
\end{align}
where $k = k_0 \to T -1$ represents the duration of the simulation time from the initial ($k_0$) to the final simulation time (T), $j$ is the sampling time-step over $N_p$. Note that in this paper, the prediction model fully utilizes the nonlinearity of the MG in~\eqref{dynmvdc} to allow the controller to predict the system's future behavior accurately.

In every time step $k$, predictions of the future behavior of the system and the optimal control sequence are solved over $N_p$. In this work, the optimal control algorithm is centered on ensuring a stable voltage and output power trajectory during mission. Therefore, a reference tracking objective function while simultaneously minimizing control input changes with guaranteed stability and recursive feasibility is formulated:
\begin{align}
    J(\bm{x},\bm{u}, \epsilon)|k &= \sum_{j = 0}^{N_p -1} J_s(\bm{x}(k+j)|k,\bm{u}(k+j)|k) \\
    &+ J_f(\bm{x}(k+N_p)|k) + J_{\epsilon}(\epsilon(k)|k)\label{mod-obj}
\end{align}
where $\epsilon(k)$ is the nonnegative slack variable at interval $k$, $\bm{x}(k+N_p)|k$ is the state at time $k+N_p$, predicted at time instant $k$. Each objective function ($J|k$) is defined by:
% \begin{subequations}
    \begin{align} \label{costcombined}
    \begin{cases}
    J_s(k+j)|k &= \|\bm{x}(k+j)-\bm{x}^{ref}(k+j)|k\|^2_{\mathbf{W_x}_j}  
    \\ &+ \|\bm{\Delta u}(k+j)|k\|^2_{\mathbf{W_{\Delta u}}_j} \\
    J_f(k+N_p)|k &= \|\bm{x}(k+N_p)-\bm{x}^{ref}(k+N_p)|k\|^2_{\mathbf{W_P}} \\
J_{\epsilon}|k &= \rho_{\epsilon}\epsilon^2(k)
    \end{cases}
\end{align}
% \end{subequations}
where $\|.\|$ denotes the Euclidean norm, $J_{\epsilon}|k$ is the slack cost function with $\rho_{\epsilon}$ as the constraint violation penalty weight, $J_s(.)$ is the stage cost with $\mathbf{W_x} \in \mathbb{R}^{n_x \times n_x} \succ 0$, and $\mathbf{W_{\Delta \hat{u}}} \in \mathbb{R}^{n_u \times n_u} \succ 0$ are the positive definite weighting matrices on the states, and the control input rate, respectively, and $\bm{x}^{ref} \in \mathbb{R}^{n_x}$ is the vector of desired set-points. To guarantee the closed-loop stability, a terminal cost, denoted as $J_f(.)$ is added with $\mathbf{W_P}\in \mathbb{R}^{n_x \times n_x} \succ 0$ as the terminal penalty weight. In order to compute $\mathbf{W_P}$, Jacobian linearization of the system is first computed around the neighborhood of the equilibrium point in \eqref{ssmvdc} as $\bm{x}_e, \bm{u}_e$, as follows:
\vspace{-0.1cm}
\begin{equation}
\vspace{-0.1cm}
    \bm{\hat{x}}(k+1) = \mathbf{A}_d \bm{x}(k) + \mathbf{B}_d \bm{u}(k) \label{linerrdyn}
\end{equation}
where $\mathbf{A}(i,j) = \frac{\partial \mathbf{f}_i}{\bm{x}_j} \in \mathbb{R}^{n_x \times n_x}$ and $\mathbf{B}(i,j) = \frac{\partial \mathbf{f}_i}{\bm{u}_j}\in \mathbb{R}^{n_x \times n_u}$ are the system's matrices and assumed to be stabilizable. Accordingly, there exists a linear state feedback control $\bm{u} = \mathcal{K}\bm{x}$ such that $\tilde{\mathbf{A}} = \mathbf{A} + \mathbf{B} \mathcal{K}$ is asymptotically stable \cite{fang2022model}. The terminal penalty weight is then can be derived by solving the Lyapunov discrete equation:
\vspace{-0.1cm}
\begin{equation}
\vspace{-0.1cm} \label{lyap}
    (\mathbf{A} + \mathbf{B} \mathcal{K})\mathbf{W_P}(\mathbf{A} + \mathbf{B} \mathcal{K})^T - \mathbf{W_P} + \mathbf{W_x}^{\ast} = 0
\end{equation}
where $\mathbf{W_x}^{\ast} \coloneqq \mathbf{W_x} + \mathcal{K}^T\mathbf{W_{\Delta u}}\mathcal{K}$. Given that $\mathbf{W_x}$ is selected to be positive definite, then  $\mathbf{W_P}$ is the unique positive-definite. 

To guarantee the recursive feasibility of the MPC optimal control problem formulation, terminal inequality constraints are added as follows:
\begin{align}\label{terminalconst}
    \begin{cases}
    \bm{x}(N_p):= \mathbf{f}_d(\bm{x}, \mathcal{K}(\bm{x}), \bm{d}) \leq \epsilon \in \mathbb{X}_f  \quad \mathbb{X}_f \subseteq \mathbb{X}\\
    \epsilon \geq 0
    \end{cases}
\end{align}
Assuming that there exists a local stabilizing controller $\mathcal{K}(\bm{x})$ for which the terminal set $\mathbb{X}_f$ is invariant (once trajectory of $\bm{x}$ enters $\mathbb{X}_f$, it will never leave the set), constraint \eqref{terminalconst} holds.

Considering the terminal ingredients and objective functions in \eqref{costcombined} and \eqref{terminalconst},  the system's dynamics and constraints \eqref{dynmvdc}, the optimal control problem formulation of LNMPC is expressed as follows:
\begin{subequations} \label{eq:tocp}
\begin{align}
    \min_{\mathbf{U},\epsilon} & \quad  J(\bm{x},\bm{u},\epsilon)|k \label{objfun}\\
    \text{s.t.} \quad & \bm{x}(k+j+1) = \mathbf{f}(\bm{x}(k+j), \bm{u}(k+j)), \label{dynamic}\\
    & \quad \quad \quad \quad \quad  j = 0, 1, \dots, N_p-1 \nonumber \\ 
    & \bm{x}(0) = \bm{x}_{plant}(k) \label{plant} \\
    & \bm{u}(k+j) \in \mathbb{U}, \quad j = 0, 1, \dots, N_p-1 \label{ubound}\\ 
    & \bm{x}(k+j) \in \mathbb{X}, \quad j = 1, 2, \dots, N_p \label{statebound}\\
    & \bm{x}(N_p) - \epsilon \in \mathbb{X}_f \subseteq \mathbb{X} \label{terminalset} \\
    &\epsilon(k) \geq 0
\end{align}
\end{subequations}
where $\mathbf{U} = [{\bm{u}(k|k),\hdots,\bm{u}{(N_p-1|k)}}]$ is the sequence of the control input, $k:= k\delta t \in \mathbb{R}^+$ is the current time step at sampling time $\delta t$ with $\bm{\Delta u}(k) \in \mathbb{R}^{n_u}$ denotes the input's rate of change vector at time step $k$. Constraints \eqref{dynamic}- \eqref{plant} denote the predicted trajectory of the navy MVDC shipboard MG evolving from $\bm{x}(0)$. Constraints \eqref{ubound}-\eqref{statebound} are the hard constraints of the system states and inputs with $\mathbb{X}:=\{\bm{\hat{x}} \in \mathbb{R}^{n_x}| A_xx\leq b_x\}$ and $\mathbb{U}:=\{\bm{\hat{u}}\in \mathbb{R}^{n_u}| A_uu\leq b_u\}$ are convex sets, and constraint \eqref{terminalset} as the soft terminal inequality constraints in \eqref{terminalconst}.
\begin{algorithm}[t!]
            \caption{Lyapunov-based nonlinear MPC (LNMPC)}
            \label{alg:lnmpc}
            \begin{algorithmic}[1]
            \State \textbf{Offline:}
            \State \quad Input: $N_p,T,n_x,n_u, \bm{x}^{ref}, \mathbf{d} \in \mathbb{D}$
            \State \quad Define $\mathbf{W_x},\mathbf{W_{\Delta u}}$, $\mathbf{f}_d(\bm{x}, \bm{u}, \bm{d}) $
            \State \quad  Linearize \eqref{ssmvdc}, \textbf{return} $\mathbf{A}_d$ and  $\mathbf{B}_d$
            \State \quad \textbf{Solve} \eqref{lyap}, \textbf{return} $\mathbf{W_P}$
            \State \quad Compute terminal constraint \eqref{terminalconst}, $\mathbb{X}_f$
            \State \quad Initialize $\bm{x} \in \mathbb{R}^{n_x},\bm{u}_0 \in \mathbb{R}^{n_u},k \leftarrow 0$
            \State \textbf{Online}
            \For {$\mathrm{k=0 \to T-1}$} \Comment{Simulation time}
            \For {$j=0$ to $N_p-1$}
           \State Develop prediction model,~\eqref{nmpc_pred} $\leftarrow$ \eqref{ssmvdc} 
            \State Formulate $J(\bm{x}, \bm{u},\epsilon)|k$ from \eqref{eq:tocp}
            \EndFor
            \State Solve $J|k$, \textbf{return} $[{\bm{u}^{\ast}(k|k),..,\bm{u}^{\ast}{(N_p-1|k)}, \epsilon|k}]$.
            \State Apply $\bm{u}(k) = \bm{u}^{\ast}(k|k)$ to \eqref{ssmvdc}, \textbf{return} $\bm{x}(k+1|k)$
            \State Update for $k+1$, $\bm{x}_0 = \bm{x}(k+1|k)$ and $\bm{u}_0 = \bm{u}^{\ast}(k|k)$ 
            \EndFor
            \end{algorithmic}
        \end{algorithm}
\section{Case Studies} \label{cs}
This section demonstrates a series of case studies to assess Lyapunov-based NMPC (LNMPC), highlighting the stability benefits for MVDC naval shipboard MGs. A detailed physical model of the power system is developed in a Simulink environment, replicating the real-world system, to validate the effectiveness of the proposed method. The proposed control algorithm is solved in MATLAB using ``fmincon" nonlinear solver with a solution method specified to sequential quadratic programming (SQP). The models are carried out on an Intel Core CPU i7-6700 processor at 3.40 GHz and 32GB RAM. The studies include: 1) comparing LNMPC's performance against a PI controller under time-varying loads ($P_{PPL}$ and $P_{CPL}$), and 2) examining the control's robustness against stressful loads including white noise disturbances. To quantify the effectiveness of the proposed method, mean absolute percentage error (MAPE) is used and expressed as follows: 
\begin{equation}
    MAPE = \frac{1}{n} \sum_{i=1}^{n} \left| \frac{{x_i - \hat{x}_i}}{{x_i}} \right| \times 100\%  \label{mapeeq}
\end{equation}
where, $n$ represents the total number of samples, $x_i$ represents the desired value, and $\hat{x}_i$ represents the predicted value.
% \begin{figure}[t!]
%     \vspace{-0.5cm}
%     \centering
%     \includegraphics[width =1\columnwidth]{Figs/pgencrp.pdf}
%     \caption{CS-1: Power balance under PPL operation}
%     \vspace{-0.5cm}
%     \label{fig:cs1pgen}
% \end{figure}

%%%%%%%%%%%%%%%%%%%%%%%%%%%%%%%%%%%%%%%%%%%%%%%%%%%%%%%%%
\subsection{Case study 1: Comparison of LNMPC and PI controllers}
This case compares the performance of LNMPC with the state-of-art control, PI controller in maintaining bus voltage subject to constant and pulsed load operation. To validate the effectiveness, a constant load ($P_{CPL}$) is applied to the system at 10MW while the pulsed loads ($P_{PPL}$) are initiated at t~=~2s with 1s duration at 3MW and repeated at t = 5s with a higher magnitude at 5MW with a duration of 2s. 

\begin{figure}[ht!]
% \vspace{-0.5cm}
\centering
\subfloat[CS-1: Output voltage between PI controller and LNMPC]
{\includegraphics[width=1\columnwidth]{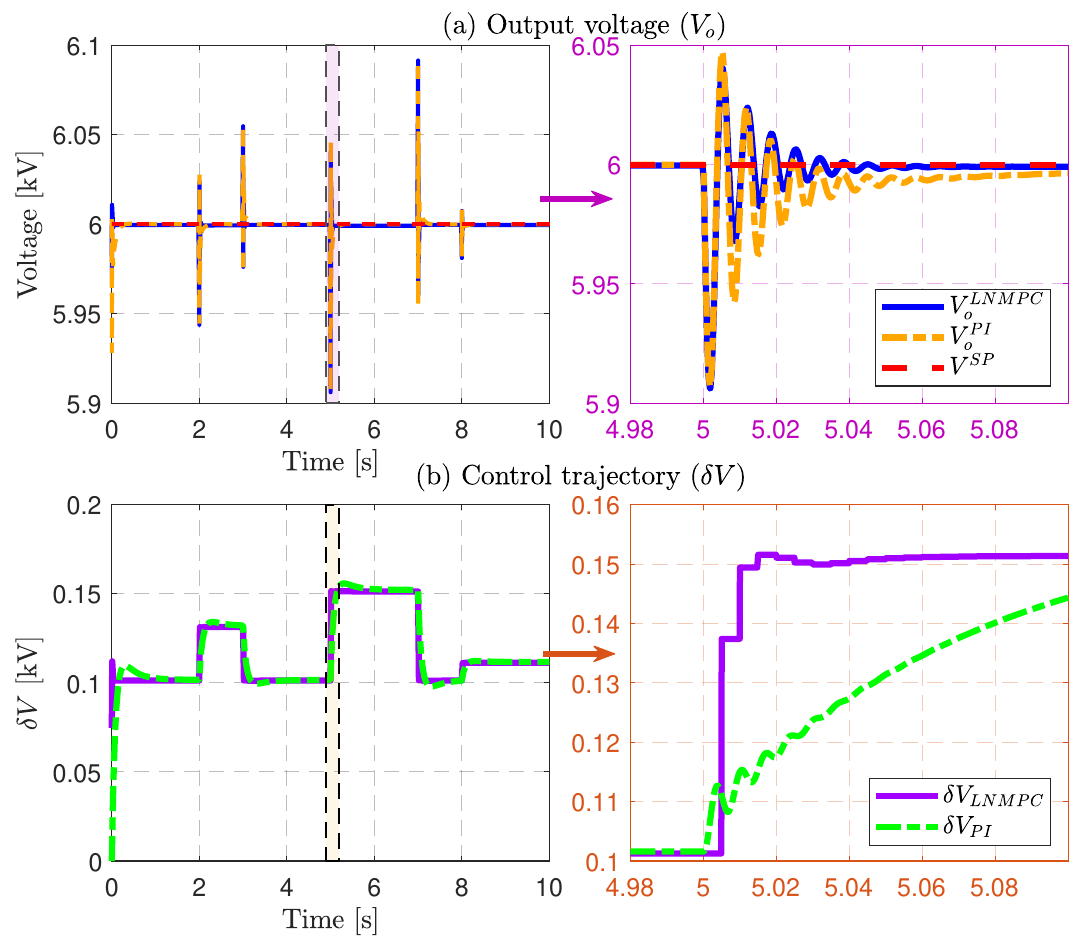}%
\label{cs1volt}}
\hfil
\subfloat[CS-1: Power sharing comparison of the distributed generation units.]
{\includegraphics[width=1\columnwidth]{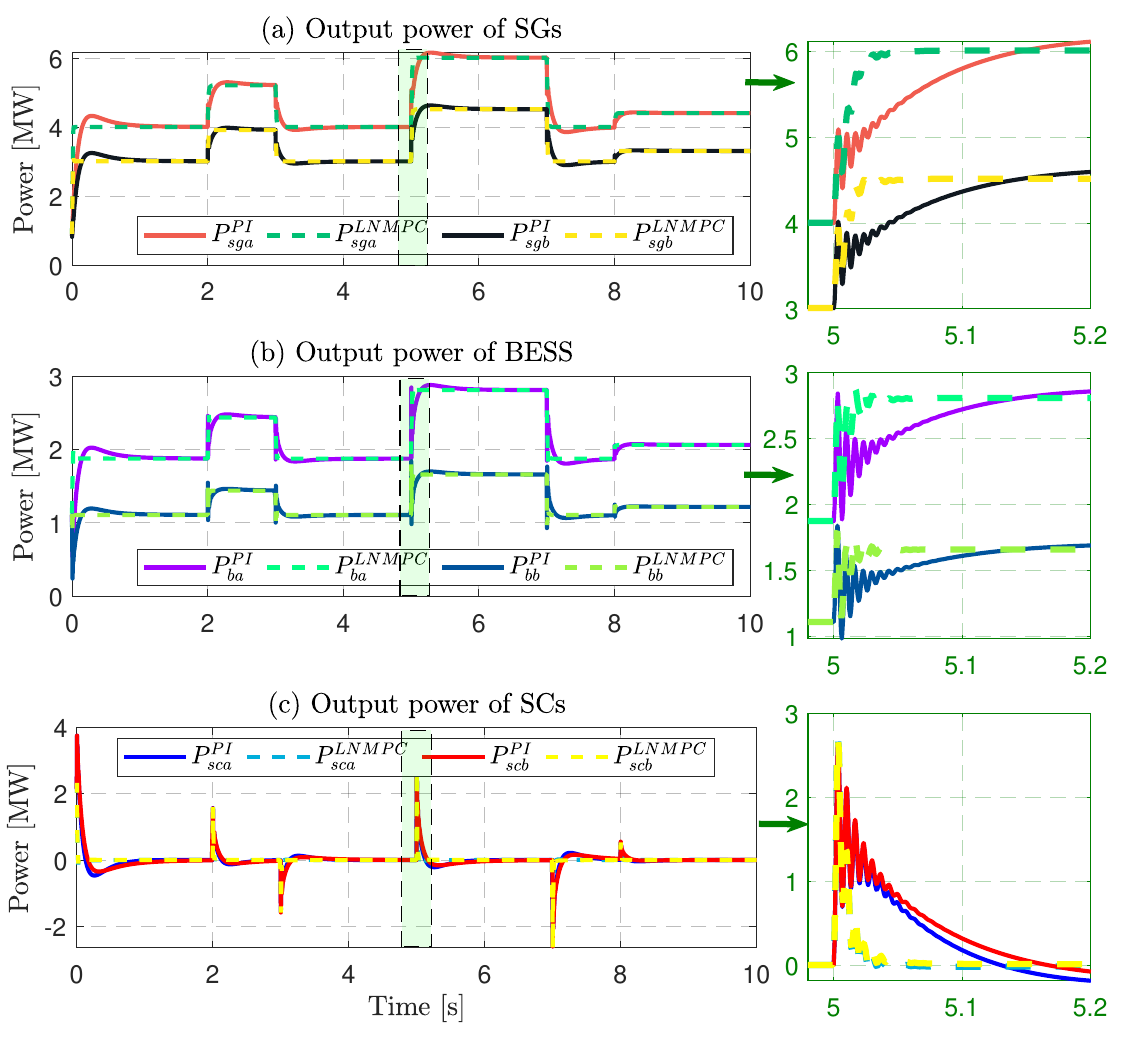}%
\label{cs1pshare}}
\caption{CS-1: MVDC naval shipboard MGs via Lyapunov-based NMPC and PI controller}
\label{cs1results}
\vspace{-0.5cm}
\end{figure}

% Fig.~\ref{fig:cs1pgen} shows the power consumption of the system in comparison to the cumulative power generated by the distributed generation units (DGUs). It can be observed that throughout the 10s simulation, the power balance is consistently met, as indicated by the overlapping line between the dispatched power and the load. 
Fig.~\ref{cs1volt} illustrates the output voltage ($V_o$) of the proposed MVDC naval shipboard MGs. For both controllers, however, $V_o$ quickly restores to the reference point with voltage variation within regulated limits at $\pm 5\%$, specifically only up to $ 1.67\%$ variation. Additionally, as observed from the transient state behavior, LNMPC outperforms the PI controller with a 50\% shorter settling time with MAPE at 0.007\%. This is due to additional stabilizing ingredients in the terminals that ensure the output voltage will return to the nominal value. 
Furthermore, Fig.~\ref{cs1pshare} shows the power generation trajectories of the DGUs under both PI controller and LNMPC.  According to the transient behavior, LNMPC managed to stabilize the output power of each DGU four times faster than the PI controller, confirming the efficacy of the proposed LNMPC for navy SPS. Furthermore, the incorporation of capacitive droop gains into the dynamics of the SCs contributed significantly to their output power, where they consistently respond to high-frequency power and avert considerable bus voltage sag. Under the LNMPC configuration, SCs are rapidly adjusted to 0MW during steady state conditions within 0.05s, whereas the PI controller requires a 0.2s longer time frame.

\subsection{Case study 2: Robustness of LNMPC subject to noise-influenced and intense load variation}
Building upon the effective performance of LNMPC when compared to a PI controller, this section evaluates the robustness of the proposed LNMPC subject to load uncertainty from $P_{CPL}$ and $P_{PPL}$. This is done by integrating amplified white noise into the loads. Furthermore, a step change to load reduction of the $P_{CPL}$ is initiated at t~=~8s while a magnitude- and time-varying $P_{PPL}$ is initiated to the system at t~=~2, 5, and 7s with 1s and 2s duration to further stress the system.  

\begin{figure}[ht!]
 \vspace{-0.5cm}
\centering
\subfloat[CS-2: Impact of load uncertainty on the output voltage]
{\includegraphics[width=\columnwidth]{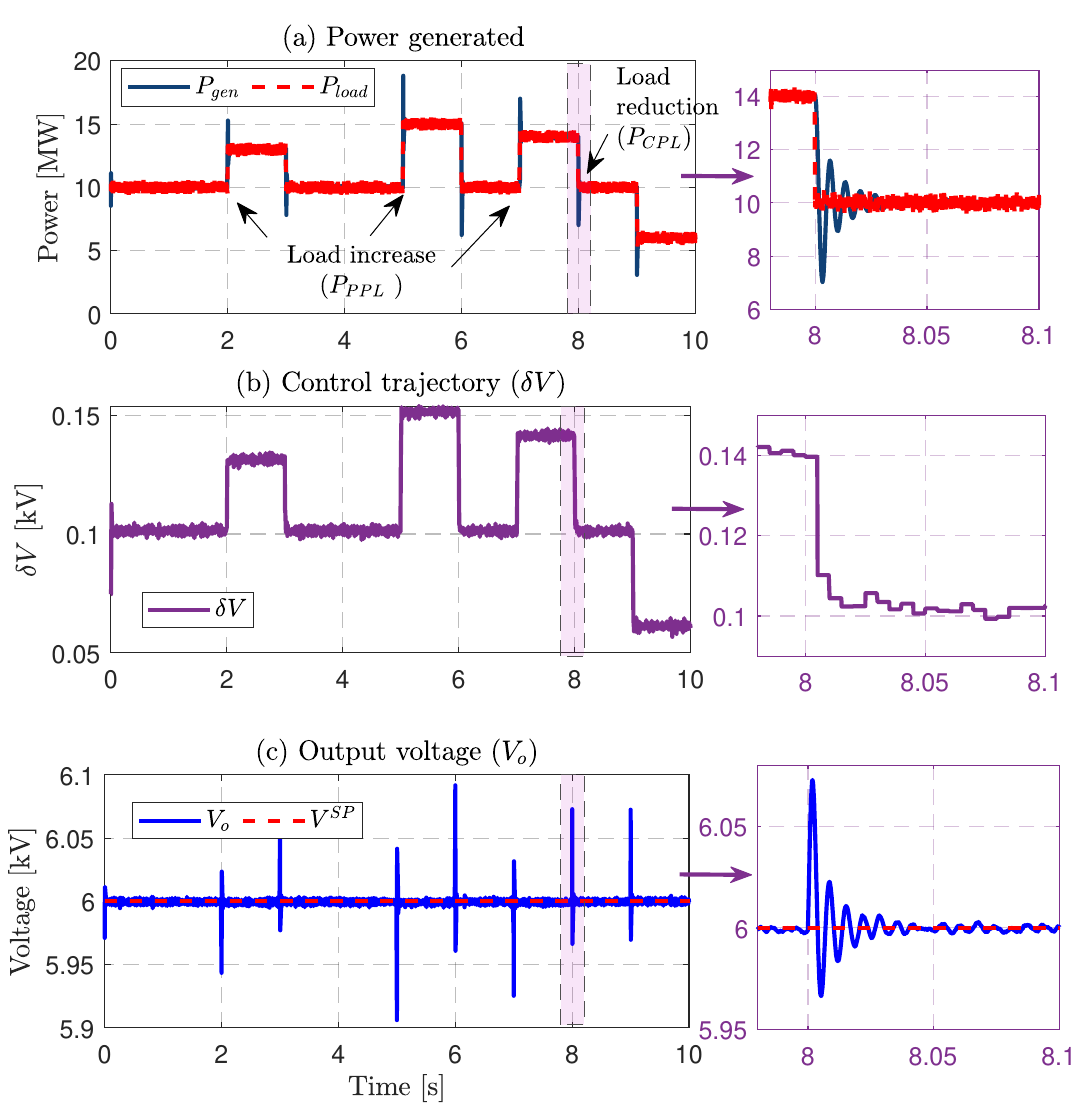}%
\label{fig:cs2volt}}
\hfil
\subfloat[CS-2: Power sharing between the DGUs]
{\includegraphics[width=\columnwidth]{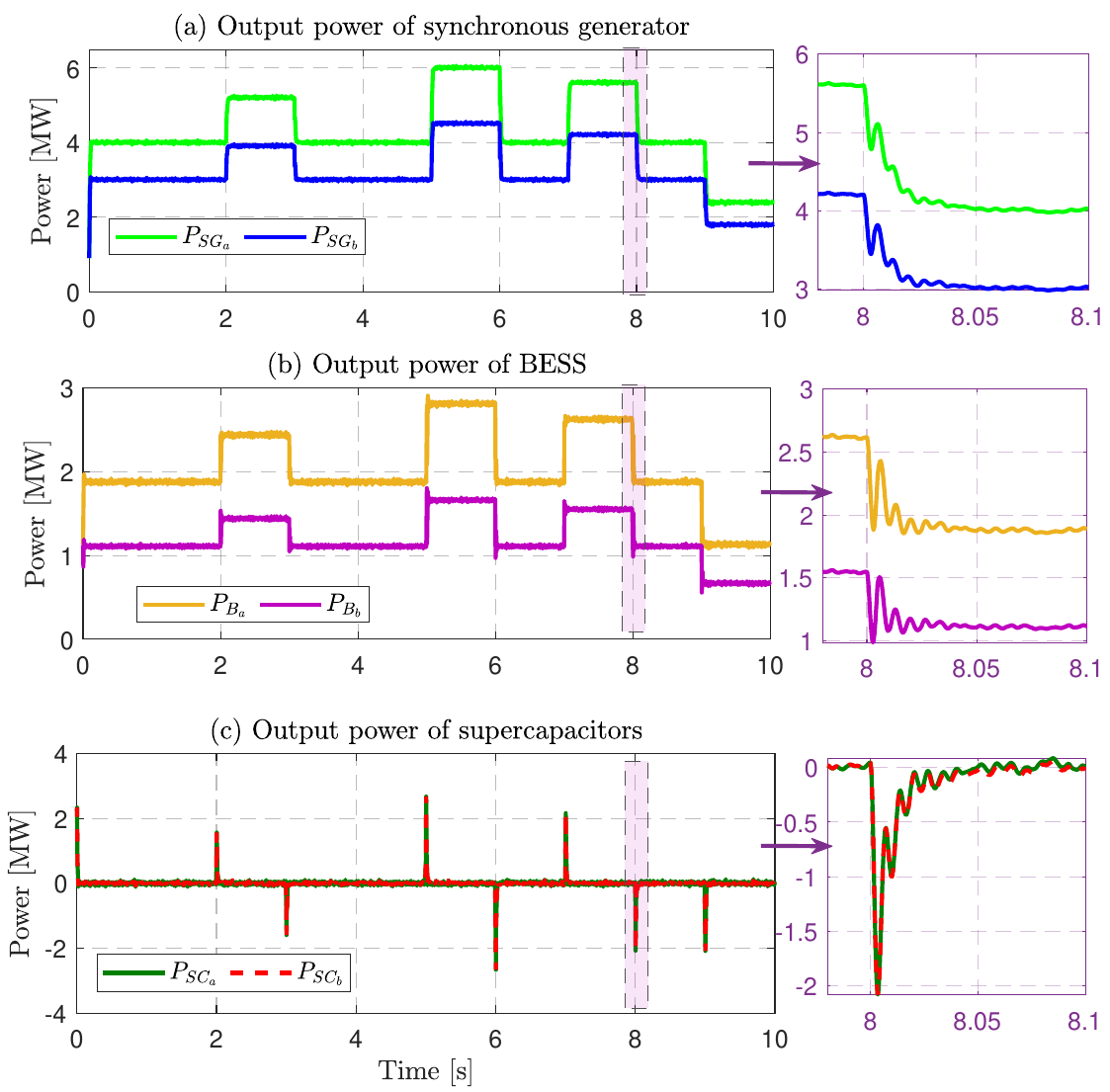}%
\label{fig:cs2pgen}}
\caption{CS-2: Lyapunov-based NMPC for MVDC naval shipboard MGs under load uncertainty}
\label{fig_cs2}
\vspace{-0.5cm}
\end{figure}

Fig.~\ref{fig:cs2volt} shows the (a) cumulative power generation from DGUs compared to the summation of $P_{PPL}$ and $P_{CPL}$, (b) $\delta_V$ trajectory to ensure constant $V_o$ and (c) output voltage trajectory of the proposed navy shipboard MG. In spite of the presence of the noisy, varying, voltage-sensitive, and high-frequency power requirement, LNMPC continuously maintains the power balance and bus voltage within the permissible boundaries ($\pm 5\%$ differences) with a tracking error of 0.02\%. The auxiliary voltage restoration signal, $\delta_V$, managed to quickly adapt to load changes to restore the output voltage to the nominal value.

Fig.~\ref{fig:cs2pgen} shows the load sharing among the DGUs (SGs, BESS, and SCs). Similar to case study 1, with the aid of the virtual resistive and capacitive droop control integrated with LNMPC, power sharing with stable trajectories is still obtained despite the presence of uncertain load patterns. It is consistent with the designed droop controller where $P_{SGa}$ contributes the highest power while $P_{Bb}$ is the lowest and $P_{SCi}$ is only utilized during transient conditions. As observed, all output power reached a steady state within 0.05s. Accordingly, LNMPC exhibits satisfactory performance under uncertainty load, which is crucial for real-world applications where ideal, noise-free conditions are infrequent.
\section{Conclusion}
This paper introduces an integrated primary and secondary control system for MVDC naval shipboard MGs, combining virtual resistor and capacitor droop control for optimal power sharing, and nonlinear MPC with stabilizing terminal cost and terminal inequality constraints for nominal bus voltage restoration. 
These components not only restore voltage but also ensure closed-loop stability and recursive feasibility through the Lyapunov equation. Performance validation involves two case studies: 1) comparing voltage and power management under time-varying pulsed loads with a PI controller, and 2) assessing the robustness of the proposed control under noise-influenced and stressful loads.
It was shown that the LNMPC maintains power demand adherence and achieved the bus voltage regulation with MAPE of 0.007\% and outperforms the PI controller in steady-state behavior, reaching stability twice as fast for voltage and output power. Furthermore, LNMPC effectively restores bus voltage to 6kV and manages load sharing among all DGUs within 0.05s, subject to load uncertainty.
Future research will explore machine learning strategies to better understand MVDC naval shipboard MGs dynamics, integrated with guaranteed stability and recursive feasibility nonlinear MPC. 

\bibliographystyle{IEEEtran}
\bibliography{IEEEabrv,main}
% % (used to reserve space for the reference number labels box)

% %   0.5em minus 0.4em\relax Harlow, England: Addison-Wesley, 1999.
% \bibliographystyle{IEEEtran}
% \bibliography{IEEEabrv,ref}
% \end{thebibliography}

\end{document}